 \documentclass[preprint,prl,12pt, superscriptaddress]{revtex4-1}
\usepackage{dcolumn}   

\newcommand{\SM}{Supplementary Information}
\usepackage[utf8]{inputenc}
\usepackage[english]{babel}
\usepackage{natbib}

\usepackage{mathtools}

\usepackage{float}
\usepackage{graphicx}
\usepackage{url}
\usepackage[colorlinks = true,
            linkcolor = blue,
            urlcolor  = blue,
            citecolor = blue,
            anchorcolor = blue]{hyperref}
\begin{document}

\title{Origin of micron-scale propagation lengths of heat-carrying acoustic excitations in amorphous silicon
}

\author{Taeyong Kim}
\affiliation{%
Division of Engineering and Applied Science, California Institute of Technology, Pasadena, California 91125, USA
}%
\author{Jaeyun Moon}
\affiliation{%
Department of Materials Science and Engineering, University of Tennessee, Knoxville, Tennessee 37996, USA}
\affiliation{%
Shull Wollan Center—a Joint Institute for Neutron Sciences, Oak Ridge National Laboratory, Oak Ridge, Tennessee 37831, USA
}%

\author{Austin J. Minnich}
 \email{aminnich@caltech.edu}
\affiliation{%
Division of Engineering and Applied Science, California Institute of Technology, Pasadena, California 91125, USA
}%

\date{\today}

\begin{abstract}
The heat-carrying acoustic excitations of amorphous silicon are of interest because their mean free paths may approach micron scales at room temperature. Despite extensive investigation, the origin of the weak acoustic damping in the heat-carrying frequencies remains a topic of debate. Here, we report measurements of the thermal conductivity mean free path accumulation function in amorphous silicon thin films from 60 - 315 K using transient grating spectroscopy. With additional picosecond acoustics measurements and considering the known frequency-dependencies of damping mechanisms in glasses, we reconstruct the mean free paths   from $\sim 0.1-3$ THz. The mean free paths are independent of temperature and exhibit a Rayleigh scattering trend over most of this frequency range. The observed trend is inconsistent with the predictions of numerical studies based on normal mode analysis but agrees with diverse measurements on other glasses. The micron-scale MFPs in amorphous Si arise from the absence of anharmonic or two-level system damping in the sub-THz frequencies, leading to heat-carrying acoustic excitations with room-temperature damping comparable to that of other glasses at cryogenic temperatures.

\end{abstract}
\maketitle

The collective acoustic excitations of amorphous solids are of fundamental interest due to their anomalous properties compared to those of crystalline solids, including an excess heat capacity at cryogenic temperatures \cite{Malinovsky_1990, Liu_1996} and damping by two-level systems \cite{Jackle_JNS_1976, Phillips_1972,HunklingerArnold_1974, Anderson1972,Zaitlin_pssb_1975}. The dispersion and damping of acoustic excitations responsible for heat transport have been extensively explored in many glasses using experimental methods such as inelastic scattering \cite{Baldi_PRL_2010, Jaeyun_PRM_2019, Monaco_PNAS_2009,Ruffle_PRL_2006, Masciovecchio_PRL_1996, Sette_Science_1998}, tunnel junction spectroscopy \cite{Dietsche_PRL_1979}, Brillouin scattering \cite{Vacher_PRB_1980, Vacher_PRB_1976, Benassi_PRB_2005}, and picosecond acoustics \cite{ZhuandMaris_PRB_1991, Devos_PRB_2008}, among others. These studies have generally found that excitations with well-defined frequency and wave vector are supported up to $\sim 1$ THz. In vitreous silica, a relative of amorphous silicon (aSi), the attenuation exhibits several different regimes, yielding different power-law frequency dependencies. For frequencies below $\sim 600$ GHz the damping scales as $\omega^{-2}$ corresponding to anharmonic and thermally activated relaxation damping. Between 600 GHz and 1 THz, a Rayleigh scattering trend of $\omega^{-4}$ is observed, followed by a return to $\omega^{-2}$ scaling \cite{Baldi_PRL_2010}. At still higher frequencies, Kittel proposed that attenuation is independent of frequency if the wavelength is comparable to the interatomic length scale \cite{Kittel_PhysRev_1949}. Considering these different regimes, the general trend of MFP versus frequency of acoustic excitations in glasses has been presented in Fig.~7 of Ref.~\cite{FA_PRB_1986} and Fig.~3 of Ref.~\cite{Goodson_JOHT_1994}, among others \cite{Zeller_PRB_1971, Graebner_PRB_1986}.

Amorphous silicon is an anomalous glass for several reasons. First, at ultrasonic frequencies, attenuation by two-level systems is observed in vitreous silica but not in aSi, suggesting a low density of these systems in aSi \cite{Von_haumeder_PRL_1980, XLiu_PRL_1997}. Second, thermal transport measurements indicate that the thermal conductivity of aSi can be higher than those of many glasses \cite{Wieczorek_1989, Kuo_1992, Moon_IJOHMT_2002,XLiu_PRL_2009} and that heat-carrying acoustic excitations travel distances on the order of one micron at room temperature despite the atomic disorder \cite{XLiu_PRL_2009,Kwon_ACSNano_2017, Braun_PRB_2016, GangChen_PRB}. This value is far larger than few nanometer value inferred for vitreous silica at room temperature \cite{Zeller_PRB_1971, Goodson_JOHT_1994}.



Experimentally  probing the damping of acoustic excitations in this regime with small sample volume is a long-standing experimental challenge, and as a result, studies of the acoustic excitations in aSi have relied on numerical simulations based on normal mode analysis. Feldman and Allen classified the excitations in aSi as propagons, diffusons, and locons according to the qualities of the normal mode eigenvectors~\cite{Allen_PhilMag_1999}. Fabian and Allen computed anharmonic decay rates of the normal modes of aSi, predicting that they should exhibit a clear temperature dependence~\cite{Fabian_PRL_1996}. Other molecular dynamics simulations based on normal mode analysis have predicted that the MFPs decrease as $\omega^{-2}$ with increasing frequency for excitations of a few THz frequency, leading to the conclusion that they are damped by anharmonicity \cite{Larkin_PRB_2014, Voltz_AIPAdv_2016}. Although some of these predictions are reported to be consistent with experiment \cite{Zink_PRL_2006}, others are not. For instance, the predicted temperature dependence of THz excitations in Ref.~\cite{Fabian_PRL_1996} is not observed experimentally using inelastic x-ray scattering \cite{Jaeyun_PRM_2019}. In the hypersonic frequency band $\sim 100$ GHz, the measured values of attenuation are lower than those predicted by anharmonic damping \cite{Hondongwa_PRB_2011}.

An experimental approach to constrain the frequency-dependence of sub-THz phonon damping in aSi would help to resolve this discrepancy. For solids like aSi with MFPs in the micron range, transient grating (TG) is a tabletop experimental method that is capable of measuring the MFP accumulation function, or the cumulative thermal conductivity distribution versus MFP \cite{Austin_Determining}. The technique relies on observations of non-diffusive thermal transport to constrain this function. The frequency-dependent MFPs and hence the damping mechanisms can be reconstructed from these measurements in certain cases with additional data. In the particular case of aSi, the dispersion of acoustic excitations is isotropic and experimentally available \cite{Jaeyun_PRM_2019}; the low and high frequency limits of the attenuation coefficient are known from picosecond acoustics and inelastic x-ray scattering \cite{Jaeyun_PRM_2019}, respectively; and the frequency-dependencies of the acoustic damping mechanisms of glasses are known \cite{Jackle_JNS_1976}, providing ample constraints for the reconstruction.



Here, we report the application of this approach to reconstruct the MFPs of sub-THz acoustic excitations in a free-standing aSi thin film. The reconstructed MFPs are independent of temperature and exhibit a Rayleigh-type scattering trend over most of the accessible frequency range from $\sim 0.1-3$ THz. These trends are inconsistent with predictions from normal mode analysis but agree with the trends measured in other glasses. The distinguishing feature of aSi is the weak anharmonic or two-level systems (TLS) damping of sub-THz vibrations, leading to acoustic attenuation in aSi at room temperature comparable to that of vitreous silica at $\sim 1$ K.



\begin{figure}
\includegraphics[width=\textwidth,keepaspectratio]{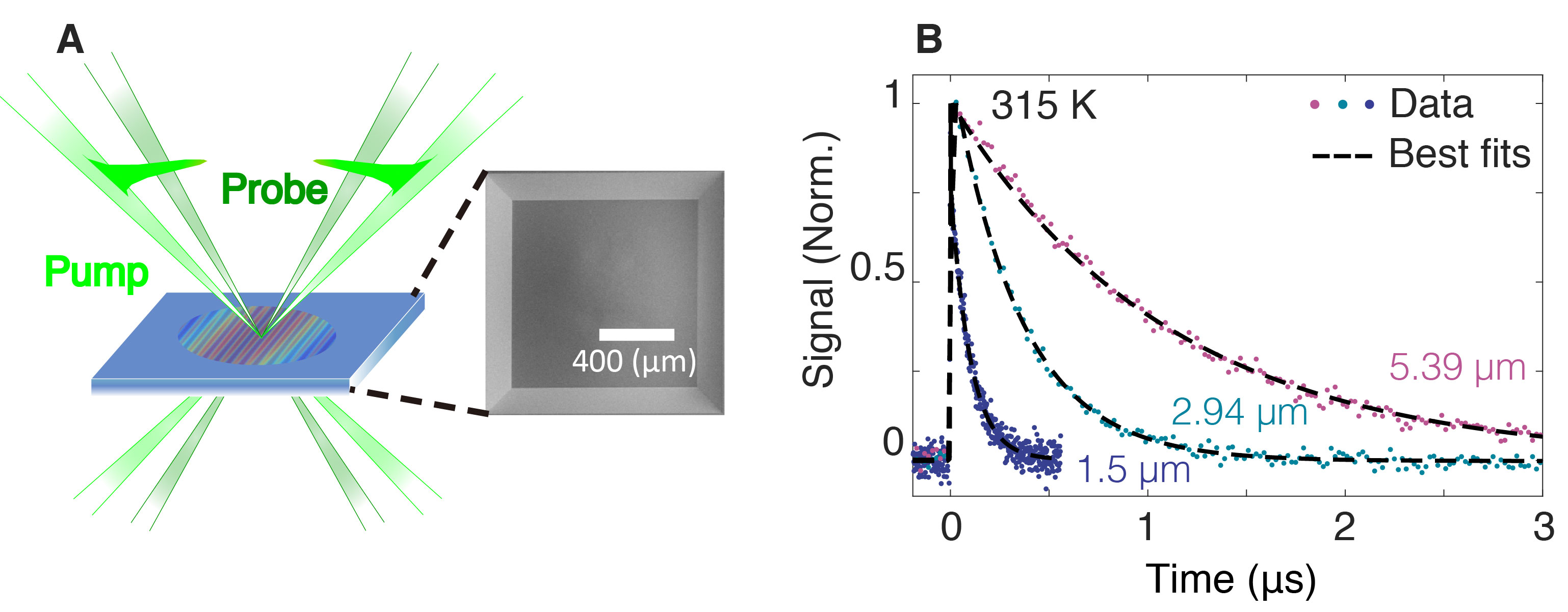}
\caption{(A) Schematic of the TG experiment. Two pump laser pulses are interfered on the sample, impulsively creating a spatially periodic temperature rise. The probe beams diffract from the transient grating, monitoring the thermal relaxation. A scanning electron microscope image of the free-standing amorphous silicon membrane is shown. (B) Representative measured signals versus time and single exponential fits for aSi at 315 K for various grating periods.}
 \label{FigaSi:Bulk}
\end{figure}

We used TG to measure the thermal diffusivity of free-standing aSi thin films with variable grating period from $\sim 0.75 -15.7$ $\mu$m. The sample of $\sim 500$ nm thickness and 1 mm$^2$ cross-sectional area is prepared by depositing aSi on a silicon nitride substrate using chemical vapor deposition as described in Ref.~\cite{Jaeyun_PRM_2019}; a scanning electron microscope (SEM) image is given in Fig.~\ref{FigaSi:Bulk}A. The transient grating setup is identical to that described in Refs.~\cite{Navaneeth_PRX, Andrew_PNAS_2019} and is illustrated schematically in Fig.~\ref{FigaSi:Bulk}A. Briefly, a pair of pump pulses (wavelength 515 nm, pulse duration $\approx$ 1 ns, repetition rate 200 Hz, 520 $\mu$m $1/e^2$ diameter) is focused on the sample to create a spatially periodic heating profile with grating period ($L$) and wave vector ($q=2\pi/L)$ defined by the incident angle. A continuous wave laser beam (wavelength 532 nm, chopped at 3.2\% duty cycle to minimize steady heating, 470 $\mu$m $1/e^2$ diameter) diffracts from the grating, monitoring its thermal relaxation. We employ a heterodyne detection method to increase the signal-to-noise ratio \cite{Maznev_OL_1998}. The optical powers were chosen to yield adequate signal-to-noise ratio while minimizing steady heating (see \SM~Sec.~V for additional discussion).



Figure~\ref{FigaSi:Bulk}B shows representative signals at $\sim$ 315 K (additional data and fits are provided in \SM~Sec.~I). The decay exhibits a single exponential profile with a time constant in the range of tens of nanoseconds to microseconds due to thermal transport. The thermal diffusivity is obtained by extracting the time constant of the single exponential decay; the thermal conductivity is then computed using the measured heat capacity from Ref.~\cite{Queen_PRL_2013}. 

\begin{figure*}[ht] 
\includegraphics[width=\textwidth,height=\textheight,keepaspectratio]{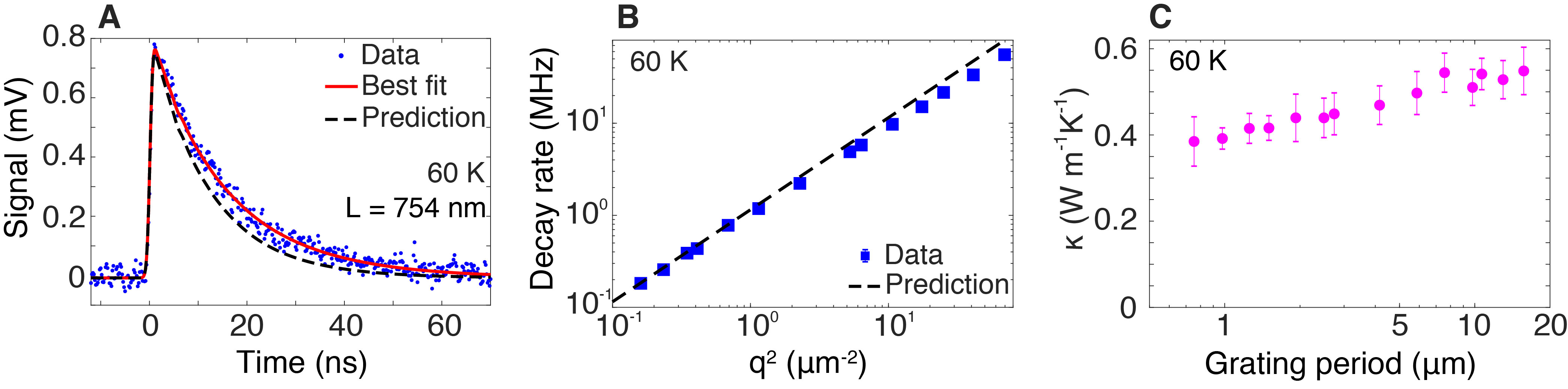}
\caption{(A) Measured TG signal versus time (symbols) for grating period $L = 754$ nm ($q^2 \sim 70$ $\mu m^{-2}$) at $T=60$ K, along with the best fit (solid red line) and the predicted decay using the thermal conductivity measured at $L=10.7$ $\mu$m (dashed black line). The actual signal decays slower than predicted, indicating a departure from diffusive thermal transport. (B) Inverse time constant versus $q^2$ at $T=60$ K. A deviation from the linear trend is observed for $L \lesssim 3$ $\mu$m ($q^2 \gtrsim 4.6$ $\mu$m). (C) Measured thermal conductivity versus grating period at $T=60$ K. A decrease in thermal conductivity of $\sim 30$\% is observed.
}
\label{FigaSi:Ballistic}
\end{figure*}

 The measured TG signal for a grating period of $L = 754$ nm is shown in Fig.~\ref{FigaSi:Ballistic}A. The actual thermal decay is clearly slower than that predicted based on the thermal diffusivity measured at $L \gtrsim 10$ $\mu$m,  indicating the presence of acoustic excitations with MFPs comparable to the grating period. Measurements of the decay rate versus $q^2$ for all the grating periods at 60 K are given in Fig.~\ref{FigaSi:Ballistic}B. The measured decay rate 
follows the linear trend expected from diffusion theory up to around $q^2 \sim 4.6$ $\mu m^{-2}$ ($L \sim 3$ $\mu$m), above which the decay rate becomes smaller.

 Figure \ref{FigaSi:Ballistic}C shows the measured thermal conductivity versus grating period obtained from these time constants at 60 K considered in this study. As the grating period becomes comparable to some MFPs, the effective thermal conductivity varies with grating period due to the ballistic transport of the acoustic excitations over the scale of the grating period. For $L = 754$ nm, the reduction of the thermal conductivity from the value at $L=10.7$ $\mu$m is $\sim$ 30\%.
%

\begin{figure}[bt] 
\includegraphics[width=.65\linewidth,keepaspectratio]{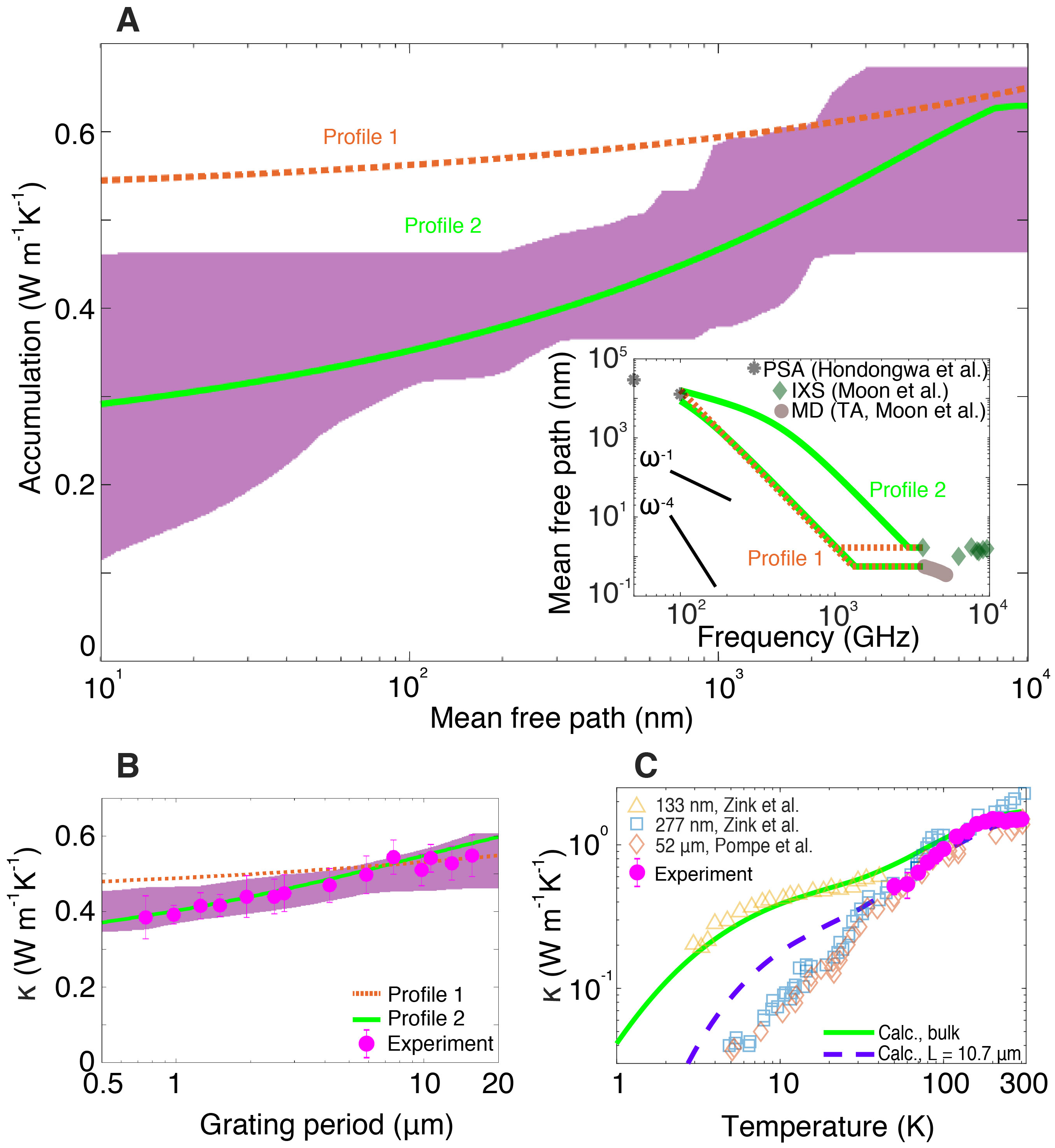} 
\caption{(A) Posterior probability distribution and 99\% credible interval of the thermal conductivity accumulation versus the mean free path at 60 K (shaded region), along with the predicted thermal conductivity accumulation at $L = 10.7$ $\mu$m from each candidate profile (lines). Inset: candidate MFP profiles versus frequency: $\omega^{-4}$, constant (profile 1: dotted lines); $\omega^{-1}, \omega^{-4}$, constant (profile 2: solid lines). Upper and lower curves indicate the LA and TA MFPs, respectively.  Literature data are presented as open symbols. (B) Measured thermal conductivity versus the grating period (symbols) along with that predicted from posterior distribution (shaded region) and the two profiles (lines) at 60 K. (C) Measured thermal conductivity versus temperature for $L = 10.7$ $\mu$m (filled circles) along with the literature data (open symbols). The calculated bulk ($L=10.7$ $\mu$m) thermal conductivities using profile 2 are shown as solid (dashed) line. 
}
\label{FigaSi:BayesMFP}
\end{figure}

We calculated the MFP accumulation function that is consistent with these measurements using the method of Ref.~\cite{Andrew_PNAS_2019}. More precisely, the posterior probability distribution of the thermal conductivity accumulation function along with credible intervals were obtained for each temperature using Bayesian inference with a Metropolis–Hastings Markov chain Monte Carlo algorithm. The posterior distribution bounded by the 99\% credible interval at 60 K is shown as the shaded region in Fig.~\ref{FigaSi:BayesMFP}A. The figure shows that the percentage of the heat carried by excitations with MFPs larger than 1 $\mu$m is $\sim$ 31 $\pm 18$\%, qualitatively agreeing with the recent observation that more than 50\% of the heat is carried by MFPs exceeding 100 nm \cite{GangChen_PRB}.

We now use the thermal conductivity accumulation function along with independent data from IXS, picosecond acoustics, and other sources to constrain the frequency-dependence of the damping. The strategy is to construct a low-energy Debye model for the thermal conductivity and identify the frequency-dependent MFPs that can simultaneously explain all of the available data. Following Ref.~\cite{Austin_Determining}, the measured thermal conductivity $\kappa_i$ can be expressed as

\begin{equation}
\kappa_{i}= \sum_s \int_{0}^{\omega_{m,s}} S(x_{i,s}) \left[ \frac{1}{3} C_{s}(\omega) v_s \Lambda_s(\omega) \right] d \omega + \kappa_{IR}(T)
\label{eq:kappa}
\end{equation}

\noindent
where $s$ indexes the polarization, $q_i = 2 \pi L_i^{-1}$, $x_{i,s} = q_i \Lambda_s (\omega) $, $S(x_{i,s})$ is the isotropic suppression function in Refs.~\cite{Austin_Determining, Maznev_PRB_2011}, $\omega_ {m,s}$ is the cutoff frequency for collective acoustic excitations, and $\kappa_{IR}(T)$ is the contribution from excitations above the Ioffe-Regel (IR) cutoff frequency. The Debye heat capacity $C_s$ of acoustic excitations is calculated from the group velocities $v_s$ which are known, isotropic, and independent of temperature \cite{Jaeyun_PRM_2019, Jaeyun_PRB_2017}. The first term of Eq.~\ref{eq:kappa} is a Debye model for the thermal conductivity of an isotropic solid that includes the effect of non-diffusive thermal transport over a grating period.

The desired quantity is $\Lambda_s(\omega)$, or the MFP versus frequency for the LA and TA polarizations. The grating period dependence of the thermal conductivity provides constraints on $\Lambda_s(\omega)$, and independent data provides further constraints. First, the linewidths of LA excitations at frequencies above 3.7 THz are known from IXS measurements and are independent of temperature \cite{Jaeyun_PRM_2019}. As the TA linewidths are not accessible with IXS, we use the values from MD simulations as these values for the LA branch quantitatively agreed with IXS measurements. These values allow the thermal conductivity of collective excitations above $\sim 3.7$ THz to be obtained from Eq.~\ref{eq:kappa}; at room temperature this contribution is $\sim 0.5$ $W m^{-1}K^{-1}$.

Second, the attenuation lengths at hypersonic frequencies $\sim 100$ GHz can be obtained from picosecond acoustics. The room temperature value is available from Ref.~\cite{Hondongwa_PRB_2011}. We performed additional measurements of the acoustic attenuation at temperatures from 30 - 300 K using this method as described in \SM~Sec.~III, and the values are on the order of 10 - 20 $\mu$m in this temperature range.

The lack of temperature dependence of the damping at both 100 GHz and $\sim 3$ THz suggests that the MFPs at intermediate frequencies should also be independent of temperature. This requirement, the PSA and IXS measurements, the measured thermal conductivity versus grating period at the 5 temperatures in this study, and the known frequency-dependencies of damping mechanisms in glasses, impose tight constraints on the frequency-dependence of the damping in the sub-THz frequencies. The inset of Fig.~\ref{FigaSi:BayesMFP}A shows two candidate MFP profiles that satisfy these constraints for the LA and TA branches. Power law dependencies $\Lambda \sim \omega^{-n}$ are assumed and combined using Matthiessen's rule. Profile 1 transitions from a constant value to $n = 4$ corresponding to Rayleigh scattering, while profile 2 transitions from constant to $n=4$ at $\sim 2-3$ THz and then to $n=1$ at $\sim 200 - 300$ GHz.

The thermal conductivity accumulation versus MFP at 60 K computed from these profiles  is given in Fig.~\ref{FigaSi:BayesMFP}A. In this figure, $\kappa_{IR}$ was chosen to match the measured thermal conductivity for $L \gtrsim 10$ $\mu$m to facilitate comparison. Profile 2 exhibits better agreement with the posterior distribution, with  profile 1 exhibiting a weaker trend with MFP compared to the posterior distribution.

The thermal conductivity versus grating period for the posterior distribution and both profiles is shown in Fig.~\ref{FigaSi:BayesMFP}B. Profile 2 again yields better agreement with the trend of thermal conductivity versus grating period compared to profile 1. An alternate profile that increases as $n=4$ immediately at 3.7 THz yields a thermal conductivity that exceeds the experimental values at all temperatures (not shown). We find that profiles consistent with profile 2 are best able to explain the magnitude and grating period dependence of the thermal conductivity. Specifically, the MFPs must remain constant as frequency decreases and then increase rapidly as $n=4$. To agree with the PSA data, the trend must then switch to $n=1$ or $n=2$. From the TG data, we are unable to determine this latter trend owing to the influence of boundary scattering in the 500 nm thick membrane; however, independent picosecond acoustic measurements at 50 GHz and 100 GHz indicate that the damping varies as $\omega^{-1}$ \cite{Hondongwa_PRB_2011}. Considering the weak dependence of the damping on temperature, this observation suggests the damping at $\sim 100$ GHz is due to density fluctuations which is predicted to exhibit an $n=1$ trend \cite{Walton_1974}.

We provide further evidence in support of profile 2 by calculating the  bulk thermal conductivity versus temperature with these MFPs and comparing it to the measured values in the present work and Refs.~\cite{Zink_PRL_2006, Pompe}. Figure~\ref{FigaSi:BayesMFP}C shows that the computed bulk thermal conductivity agrees with the measurements of Ref.~\cite{Pompe}. However, discrepancies are observed between the computed bulk thermal conductivity and the data of this work and Ref.~\cite{Zink_PRL_2006}. Accounting for the maximum grating period used in the present experiments ($L=10.7$ $\mu$m), suppressing the contribution of phonons of MFP exceeding this length scale yields good agreement with our data and qualitative agreement with Ref.~\cite{Zink_PRL_2006}. Therefore, the trends of their measurements could be attributed to extrinsic boundary scattering limiting the maximum MFP rather than intrinsic damping mechanisms. The value of specularity parameter required to produce micron-scale MFPs $\Lambda$ for the $d \sim 100$ nm thick membrane used in their work is $\sim 0.95$ using $\Lambda = d (1+p)/(1-p)$; this specularity agrees well with that measured at surfaces terminating crystalline silicon \cite{Klitsner_PRB_1987}.

Evidence in support of the $n=1$ trend at $\sim 100$ GHz can be obtained by extrapolating the MFP inferred from the dominant phonon approximation with the data of Ref.~\cite{Pompe}. The dominant heat-carrying phonon frequency at $\sim3$ K is $\sim 260$ GHz  \cite{Zeller_PRB_1971}; taking their measured thermal conductivity and the computed heat capacity of acoustic excitations at 3 K ($\kappa \sim 0.2$ $W m^{-1}K^{-1}$, and $C\sim 48$ $Jm^{-3}K^{-1}$, respectively) along with the average sound velocity ($v \sim 4400$ $ms^{-1}$), the average MFP of the dominant phonon can be obtained using $\Lambda_{dom} \sim 3 \kappa /C v \approx 3$ $\mu$m. If the MFP trend follows $n=1$ at these frequencies, the MFP from Ref.~\cite{Pompe} at 3 K implies a MFP of $\sim 10 - 20$ $\mu$m at 50 - 100 GHz, close to the PSA measurements of the present work. This analysis thus supports the magnitude, frequency dependence, and weak temperature dependence of the reconstructed MFPs.




We now reconstruct the MFP profile that best explains the data and is consistent with profile 2. Given the above constraints, the MFPs are characterized by only two parameters: the transition frequencies from $n=4$ to $n=1$ for both acoustic polarizations, $\omega_{m,L}$ and $\omega_{m,T}$. The remaining unknown parameter is $\kappa_{IR}$(T), which may depend on temperature. We let this function follow the temperature dependence of the heat capacity, $\kappa_{IR}(T) \propto C(T)$ \cite{Feldman_PRB_1993}. Then, we obtain the MFPs that best explain the TG data by numerically optimizing these parameters to fit the TG data at all temperatures. The comparison between the computed and measured thermal conductivity is provided in \SM~Sec.~II, and good agreement is observed at all temperatures. Further discussion of the choices of $\omega_{m,L}$, $\omega_{m,T}$, and $\kappa_{IR}$ that are compatible with the data is given in \SM~Sec.~IV; the trend of MFP with frequency remains the same for all of these parameter sets.

\begin{figure}[ht!] 
\includegraphics[width=\textwidth,keepaspectratio]{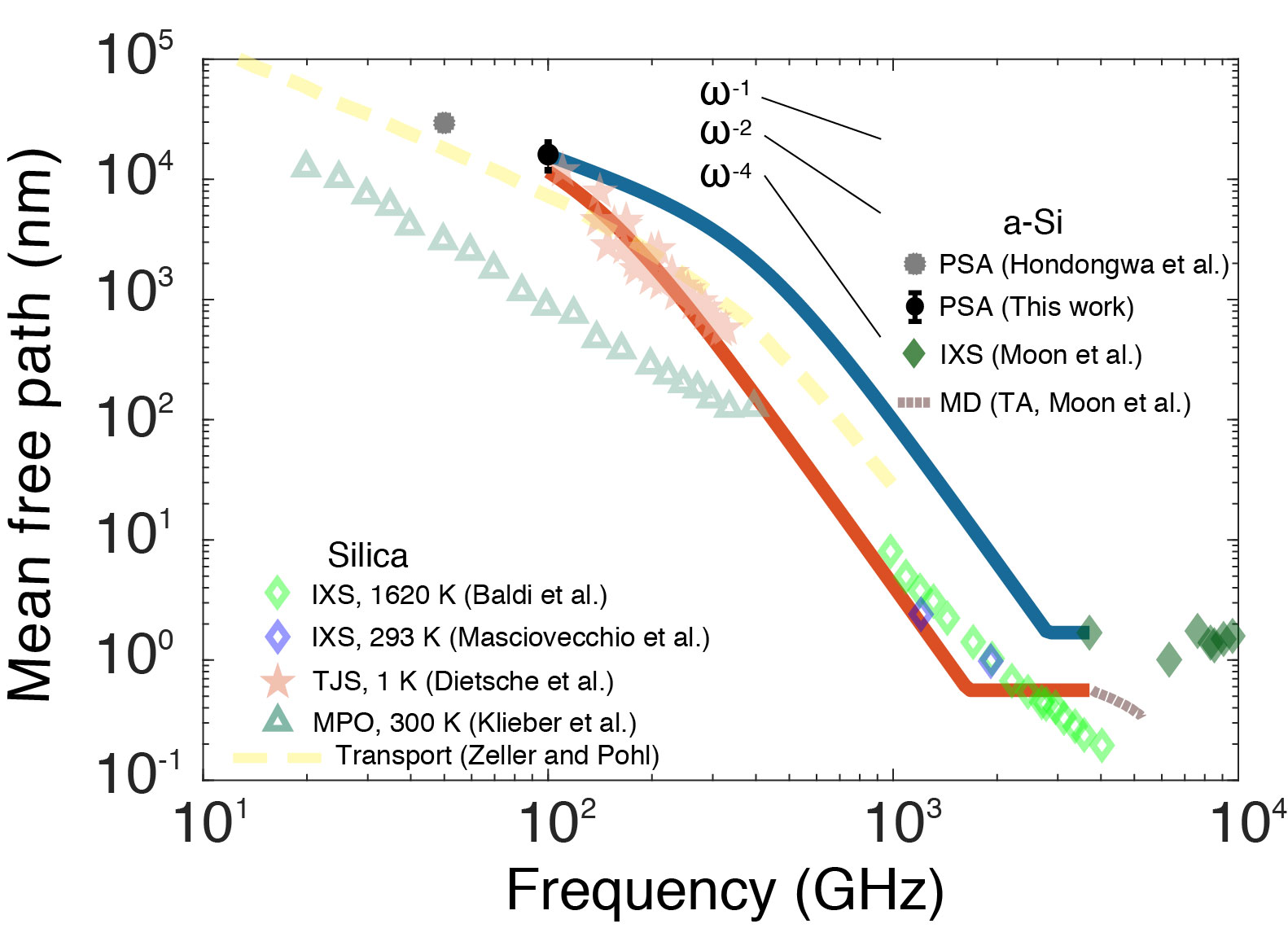} 
\caption{(A) Reconstructed mean free path versus frequency for thermal acoustic excitations in aSi at 300 K. Also shown are PSA data for aSi (Ref.~\cite{Hondongwa_PRB_2011}),  literature data for vitreous silica from inelastic x-ray scattering (diamonds, Refs.~\cite{Baldi_PRL_2010,Masciovecchio_PRB_1997}), tunnel junction spectroscopy (5-pointed stars, Ref.~\cite{Dietsche_PRL_1979}), a multi-pulse optical technique (upward pointing triangles, Ref.~\cite{Klieber_APL_2011}), and from transport measurements (dashed line, Ref.~\cite{Zeller_PRB_1971}).}
\label{FigaSi:MFPwithSilica}
\end{figure}

The reconstructed MFPs for the LA and TA polarizations are shown in Fig.~\ref{FigaSi:MFPwithSilica}. To gain physical insight into the damping mechanisms in aSi, we compare the characteristics of the reconstructed MFPs with those reported for vitreous silica, an extensively studied glass with representative measurements provided in Fig.~\ref{FigaSi:MFPwithSilica}. The frequency-dependence of the MFPs in aSi agrees well with these and other measurements \cite{Baldi_PRL_2010, ZhuandMaris_PRB_1991, Devos_PRB_2008, Vacher_PRB_1976, Benassi_PRB_2005, Dietsche_PRL_1979, Zaitlin_PRB_1975, Graebner_PRB_1986}. The $n=4$ Rayleigh scattering trend for both glasses occurs in the $1-3$ THz range with a transition to a weaker power law in the sub-THz frequencies. Thus, the frequency-dependence of acoustic damping in aSi shares several features in common with other glasses.

However, comparing the attenuation between vitreous silica and aSi,  differences emerge. First, acoustic damping in aSi exhibits little temperature dependence. Although Ref.~\cite{Zink_PRL_2006} reported a $T^{-2}$ dependence of the MFP, their measurements may have been affected by extrinsic boundary scattering as discussed above.  In contrast, the damping in vitreous silica exhibits a strong temperature dependence as observed in Fig.~\ref{FigaSi:MFPwithSilica}, highlighting the role of anharmonic and TLS damping in vitreous silica.

Other differences become clear on closer examination. Comparing the LA MFPs at $\sim 1-2$ THz, the attenuation due to Rayleigh scattering is weaker in aSi by around a factor of 5, expected as aSi is a monatomic glass with less atomic disorder. Further, at room temperature the $n=4$ trend yields to a $n=1$ trend at $\sim 700$ GHz in vitreous silica while the same transition occurs at $\sim 300-400$ GHz in aSi. At cryogenic temperatures $\sim 1$ K for vitreous silica, the transition frequencies in both materials are comparable. This difference indicates weaker damping by mechanisms such as TLS or anharmonic damping in amorphous Si and has an important consequence: excitations with MFPs in the micron range occur at frequencies of $\sim 200-1000$ GHz in aSi versus $\lesssim 100$ GHz in vitreous silica at room temperature owing to the steep $n=4$ slope of Rayleigh scattering. The heat capacity of excitations in the former frequency range is larger by a factor of $(\omega_{aSi}/\omega_{SiO_2})^2 \sim 100$. The result is that in aSi, heat-carrying excitations at room temperature have micron-scale MFPs, while the MFPs of excitations in the same frequency band for vitreous silica are smaller by an order of magnitude. Excitations in vitreous silica with micron-scale MFPs have too low frequency to transport substantial heat at room temperature. Remarkably, the attenuation observed in aSi up to room temperature is of the same order as that measured in vitreous silica at 1 K, highlighting the unusually weak acoustic damping in aSi.

Our conclusions on the origin of damping of sub-THz excitations in aSi are consistent with these prior studies of other glasses but are not consistent with the conclusions of numerical studies of excitations below the IR frequency \cite{HeGalli_APL_2011, Larkin_PRB_2014, Fabian_PRL_1996, Fabian_PRL_1999, Voltz_AIPAdv_2016}. In these studies, the Hamiltonian for atoms in a supercell is diagonalized in the harmonic approximation to yield normal modes. The original studies of Allen and Feldman used these normal modes to classify excitations in glasses as propagons, diffusons, and locons depending on the properties of the eigenvectors \cite{Feldman_PRB_1993}. The lifetimes of these modes are obtained using normal mode decomposition and molecular dynamics \cite{Ladd_PRB_1986, McGKav_PRB_2004}. With these approaches, these studies have generally concluded that damping in aSi varies as $\omega^{-2}$ for frequencies around $\sim 2$ THz and below. From this trend, the damping mechanism has been postulated to involve anharmonicity \cite{Larkin_PRB_2014, Voltz_AIPAdv_2016} and to exhibit a temperature dependence \cite{Fabian_PRL_1996}. 

We first address the classification of acoustic excitations. Various numerical \cite{Fabian_PRL_1996,Allen_PhilMag_1999, Larkin_PRB_2014} and experimental works \cite{Braun_PRB_2016} have noted a transition in the character of vibrations in aSi around $\sim 1-2$ THz, leading to the introduction of ``diffusons'' as non-propagating yet delocalized vibrations in Refs.~\cite{Fabian_PRL_1996, Allen_PhilMag_1999}. In contrast, our work attributes this transition to a change in frequency-dependent damping of collective acoustic excitations. The crossover from propagons and diffusons at $\sim 1-2$ THz coincides with the transition from Rayleigh scattering to the Kittel regime in the present work and thus can be explained without the definition of a new type of vibration. The IR crossover for the transition from collective excitations to incoherent excitations, which occurs well above 1-2 THz in amorphous Si, is sufficient to describe the different characters of excitations in glasses. 

The second inconsistency is the prediction by normal-mode analysis of the frequency-dependence ($n=2$) and anharmonic origin of damping in the few THz frequency range. Specifically, the MFPs predicted from normal mode analysis are on the order of 10 - 20 nm at $\sim 1$ THz and vary as $\omega^{-2}$ (see Fig.~4B of Ref.~\cite{Larkin_PRB_2014}), which cannot explain the measurements of the present work. In particular, extrapolating the 20 nm value at 1 THz to 100 GHz as $\omega^2$ yields  $\sim 2$ $\mu$m, which is smaller by a factor of 10 compared to the PSA measurements at the same frequency (10 - 20 $\mu$m, Ref.~\cite{Hondongwa_PRB_2011} and this work). Here, the inconsistency appears to arise from the implicit assumption of the normal mode decomposition that the heat-carrying excitations in glasses are the normal modes of the supercell. This assumption is not compatible with basic many-body physics and scattering theory, which instead gives the proper definition and lifetime of a collective excitation of a many-body system using the self-energy and the single-particle Green's function \cite{ColmanPiers_Physics}. Rather than normal modes, a physical picture of acoustic excitations of a glass that is compatible with this framework is that originally postulated by Kittel \cite{Kittel_PhysRev_1949} in the continuum limit and later by Zeller and Pohl \cite{Zeller_PRB_1971}, in which a glass is imagined to consist of a fictitious atomic lattice along with perturbations representing the mass and force constant disorder in the actual glass. The undamped excitations of the fictitious atomic lattice acquire a lifetime owing to the disorder of the actual glass. The dispersion and lifetimes of these excitations can be measured experimentally using inelastic scattering, as has been performed for many glasses in the past decades \cite{Baldi_PRL_2010, Jaeyun_PRM_2019, Monaco_PNAS_2009,Ruffle_PRL_2006, Masciovecchio_PRL_1996, Sette_Science_1998}. In contrast, the lifetimes of normal modes do not appear to be experimentally accessible or physically meaningful as they are unable to explain the thermal conductivity measurements presented here.

In summary, we report measurements of the thermal conductivity accumulation function of aSi and reconstruction of the MFPs versus frequency  using picosecond acoustics and transient grating spectroscopy. The reconstructed MFPs lack a temperature dependence and exhibit a trend characteristic of structural scattering by point defects and density fluctuations. This result is at variance with numerical studies based on normal mode analysis but is broadly consistent with prior studies of vitreous silica and other glasses. The micron-scale MFPs of heat-carrying excitations at room temperature are found to arise from the weak anharmonic and two-level system damping of sub-THz excitations, leading to room temperature attenuation coefficients comparable to those of other glasses at cryogenic temperatures.

\begin{acknowledgments}
The authors acknowledge discussions with A. B. Robbins and B. C. Daly. This work was supported by the 2018 GIST-Caltech Research Collaboration.
\end{acknowledgments}

\bibliographystyle{unsrtnat}
\bibliography{ms.bib}

\end{document}


\title{Supporting information: Origin of micron-scale propagation lengths of heat-carrying acoustic excitations in amorphous silicon
}

\author{Taeyong Kim}
\affiliation{%
Division of Engineering and Applied Science, California Institute of Technology, Pasadena, California 91125, USA
}%
\author{Jaeyun Moon}
\affiliation{%
Department of Materials Science and Engineering, University of Tennessee, Knoxville, Tennessee 37996, USA}
\affiliation{%
Shull Wollan Center—a Joint Institute for Neutron Sciences, Oak Ridge National Laboratory, Oak Ridge, Tennessee 37831, USA
}%

\author{Austin J. Minnich}
 \email{aminnich@caltech.edu}
\affiliation{%
Division of Engineering and Applied Science, California Institute of Technology, Pasadena, California 91125, USA
}%

\date{\today}
{    \global\let\newpagegood\newpage
    \global\let\newpage\relax
\maketitle}



\section{Raw Transient grating data}

Additional representative transient signals from TG are presented here.  Note that measured diffusivities in the main text were obtained by measuring multiple locations from which we determined the mean value and error bars.

\begin{figure}[h]
\includegraphics[width=\linewidth,height=\textheight,keepaspectratio]{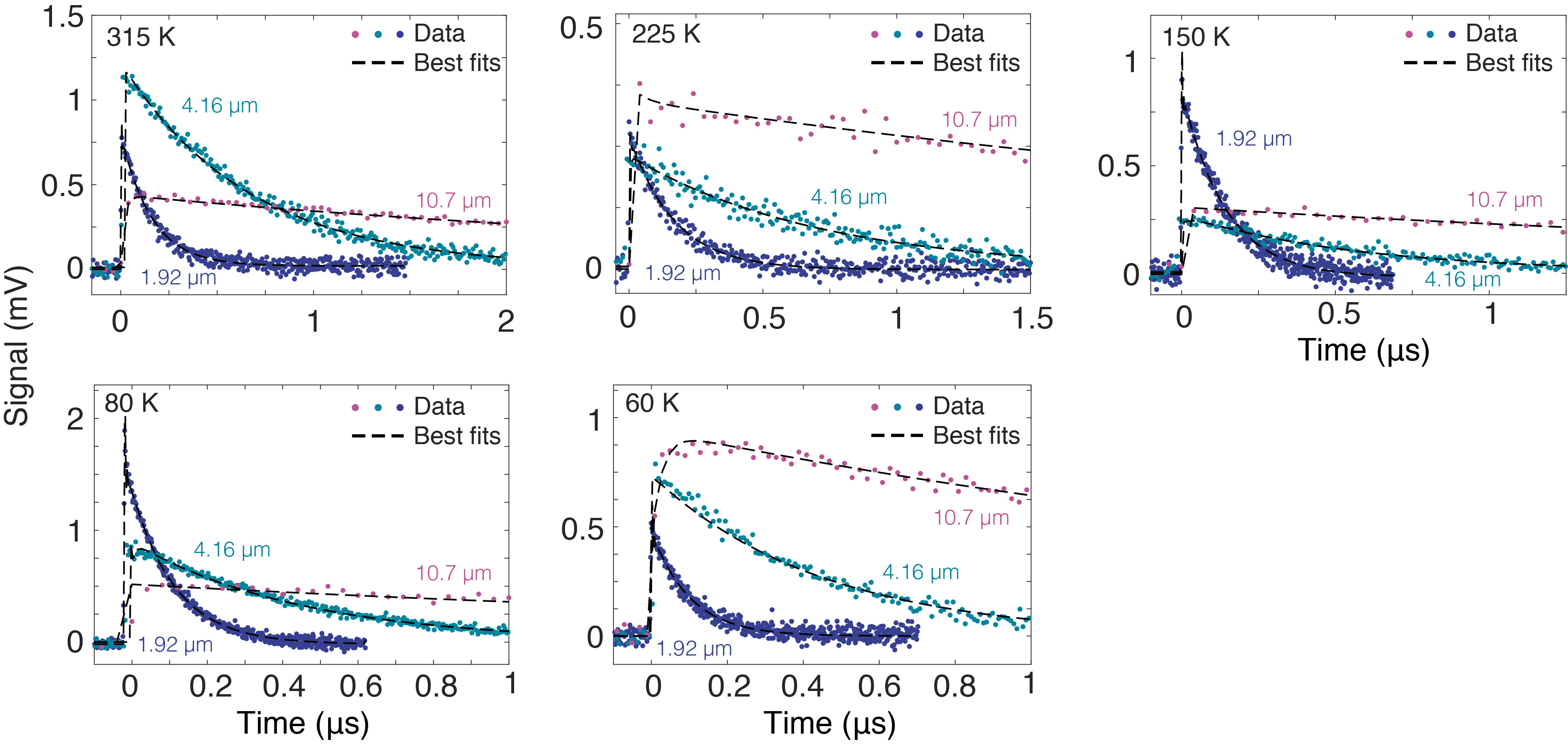}
\caption{Measured TG signal versus time from 60 K to 315 K at several grating periods.}
\label{SFigaSi:TGRawSig}
\end{figure}
\clearpage

\section{Thermal conductivity versus grating period}
\begin{figure}[!ht] \centering
\includegraphics[width=130mm,keepaspectratio]{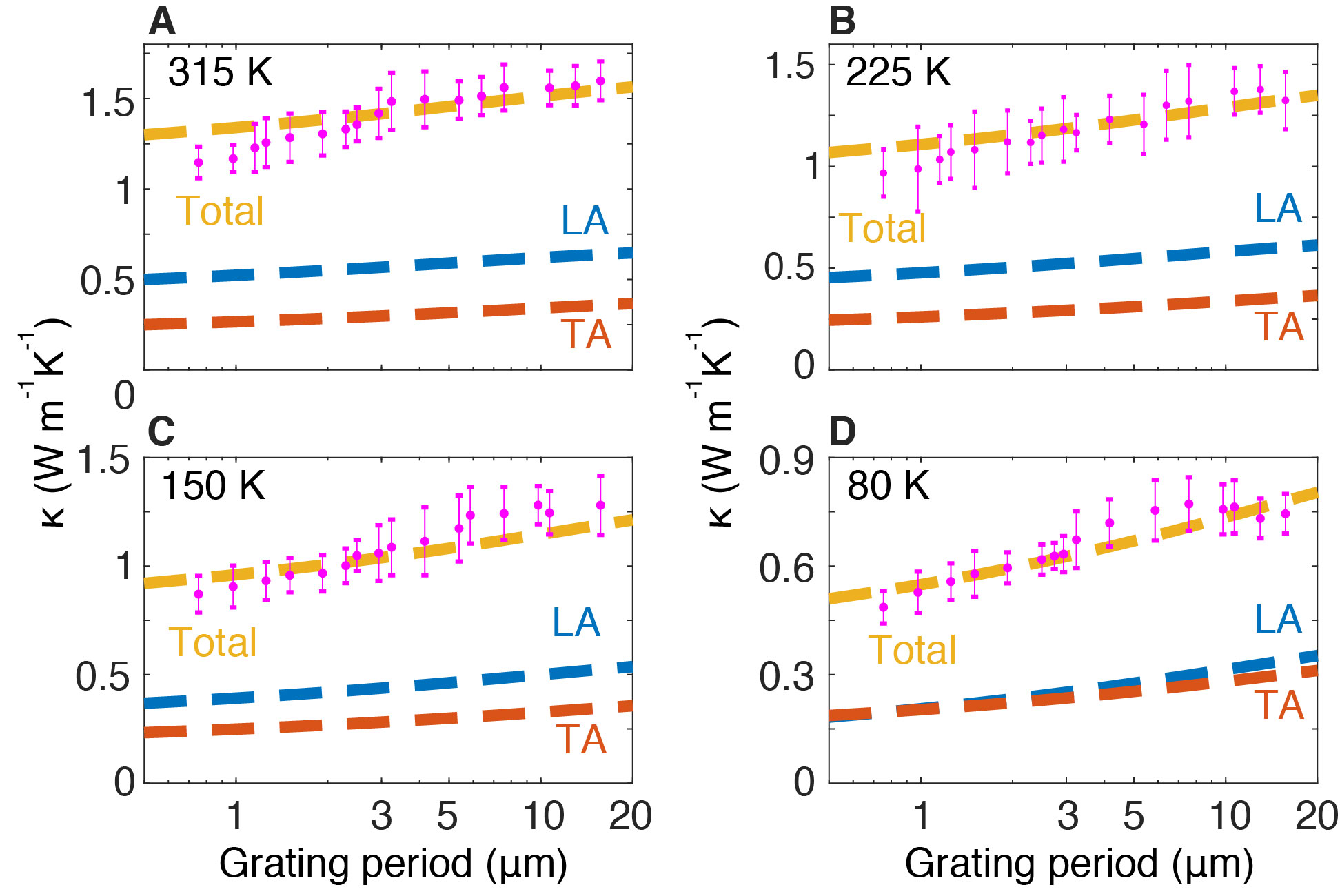}
\caption{Measured thermal conductivity versus grating period (symbols) and predicted value from optimized profile 2 (total, dashed yellow line; LA, dashed blue line; TA, dashed red line) for (A) 315 K, (B) 225 K (C) 150 K (D) 80 K. Good agreement is observed at all temperatures with MFPs that are independent of temperature.}
\label{FigaSi:fits}
\end{figure}

This section presents measured thermal conductivity versus the grating period at 315 K - 80 K, and calculation using the profile 2 discussed in the main text. Figure~\ref{FigaSi:fits} demonstrates a good agreement at all temperatures.
\clearpage

\section{Picosecond acoustics (PSA)}

The attenuation coefficient of hypersonic waves in aSi films on sapphire substrates was measured using picosecond acoustics (PSA). The samples were prepared using the same method as in the main text except with a sapphire substrate (University wafer) and a variable deposition time to prepare three samples with different thicknesses: $\sim$ 500 nm, $\sim$ 1 $\mu$m, and $\sim$ 4 $\mu$m. A 15 nm layer of Al is deposited afterwards by electron beam evaporation to serve as a transducer.

The optical system used for PSA is identical to that used in time domain thermoreflectance (TDTR). Briefly, a train of the pulses (repetition rate: 76 MHz, wavelength 785 nm) is split into pump (1/e$^2$ diameter: 13.7 $\mu$m, power: $\sim 10$ mW) and probe (1/e$^2$ diameter: 12.4 $\mu$m, power: $\sim 3$ mW) using a two-tint color method \cite{Kang_RSI_2008}. The pump is amplitude-modulated at 9.2 MHz and focused onto the sample. The pump absorption induces a thermal expansion of the Al transducer, thermoelastically launching a longitudinal strain pulse. A 15 nm Al film was chosen as a transducer to generate an acoustic of pulse around $\sim$ 100 $\pm$ 20 GHz that propagates through the aSi as in Ref.~\cite{Hondongwa_PRB_2011}. The strain pulse experiences multiple reflections between Al transducer and the sapphire substrate, producing changes in reflectance of the transducer film. Sapphire was chosen as a substrate to enhance the magnitude of the echo signals based on the acoustic mismatch model \cite{Hondongwa_PRB_2011} (for sapphire, $\rho = 3.98$ g $cm^{-3}$ , v$_{LA}$ $\sim 11000$ $m s^{-1}$) \cite{Mayer_JASA_1958}. The resulting signal is detected using a lock-in amplifier.

\begin{figure}[!hbt]
\includegraphics[width=\linewidth,height=\textheight,keepaspectratio]{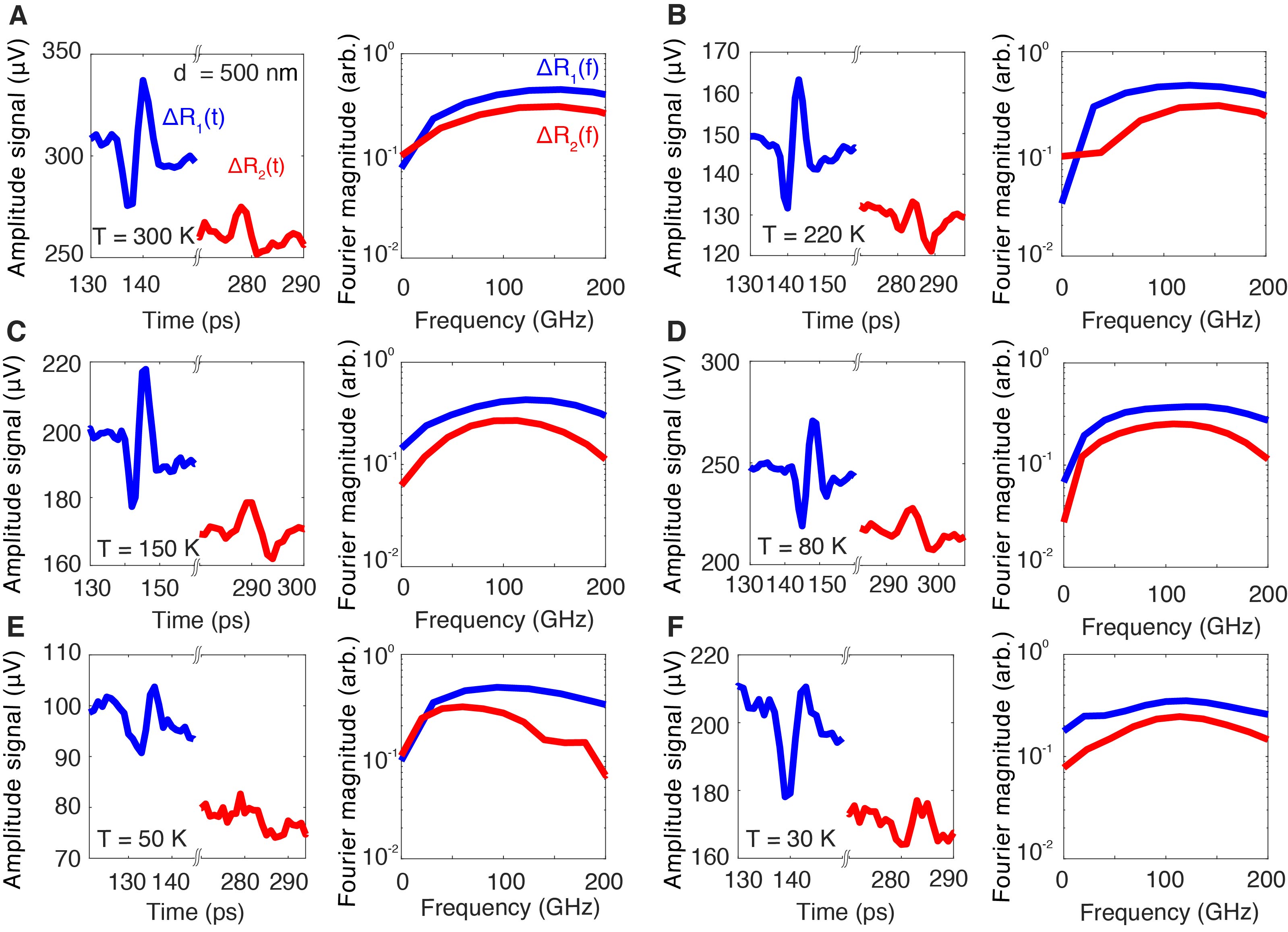}
\caption{(A) (Left) Measured amplitude signal from $\sim$ 500 nm aSi versus delay time using picosecond acoustics at 300 K. The thermal background signals have been removed. Echo 1 occurs at the Al transducer after a round trip between the transducer and the substrate. The echo 2 is the attenuated signal from echo 1 after one more round trip. The thickness of the aSi film was determined from the time difference between the echoes and the known LA sound velocity. (Right) Fast Fourier transform (FFT) magnitude of echo 1 and echo 2 at 300 K. The peak of the FFT for the first echo occurs at $\sim 100$ GHz, indicating that 15 nm Al transducer can generate acoustic waves with frequency $\sim$ 100 GHz as expected. Note the attenuation for the 500 nm film is due to the reflection or interface loss at the boundary between the aSi and the sapphire. Amplitude signal along with the corresponding magnitude of the Fourier transform measured at (B) 220 K, (C) 150 K, (D) 80 K, (E) 50 K, and (F) 30K.}
\label{SfigaSi:psaraw500}
\end{figure}

\begin{figure}[!hbt]
\includegraphics[width=\linewidth,height=\textheight,keepaspectratio]{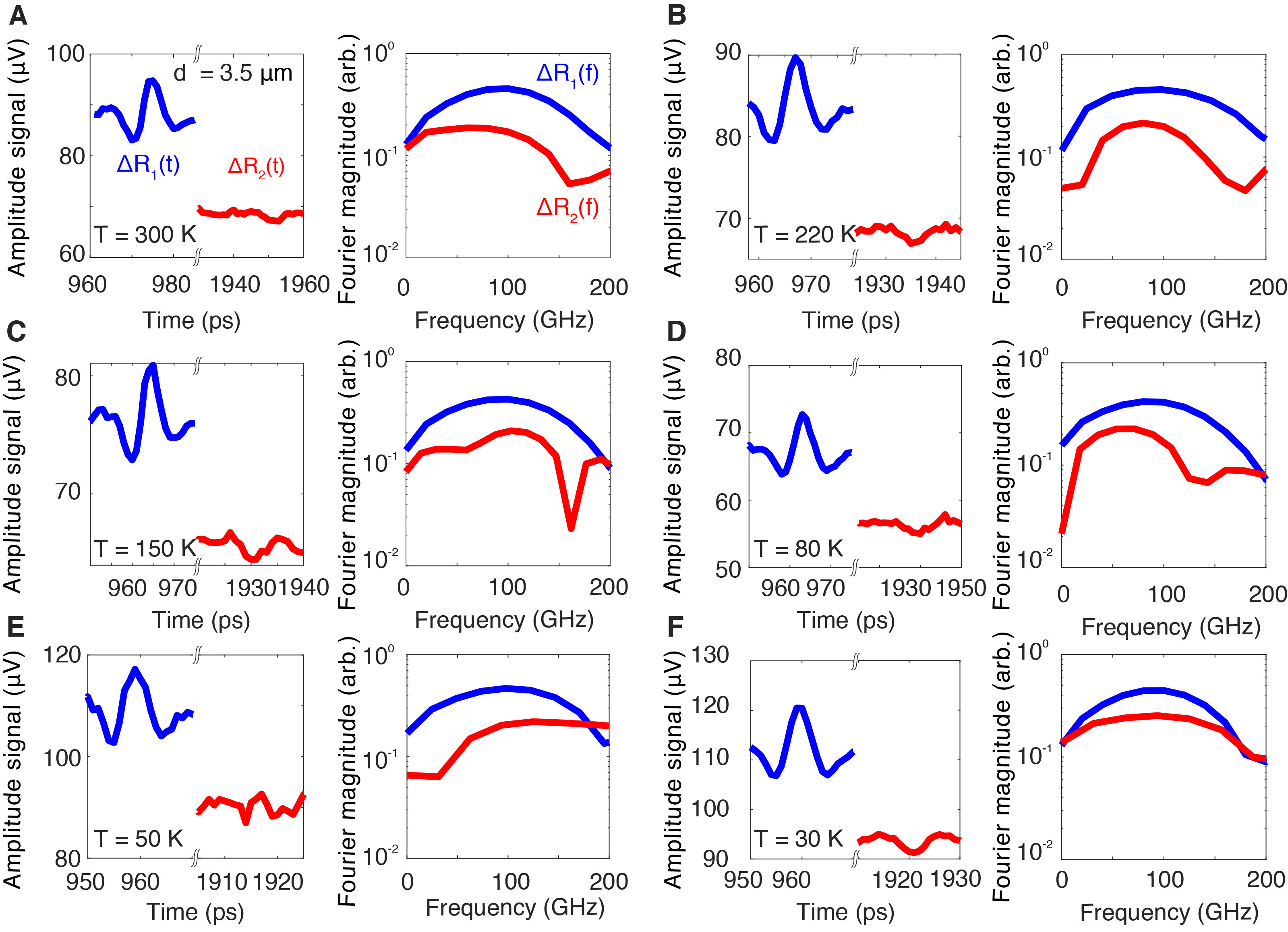}
\caption{(Left) Measured thermoreflectance signal from 4 $\mu$m sample versus delay time using picosecond acoustics at (A) 300 K, (B) 220 K, (C) 150 K, (D) 80 K, (E) 50 K, and (F) 30 K. The thermal background has been removed. (Right) Magnitude of Fourier transform of echo 1 and echo 2 at the same temperatures.}
\label{SfigaSi:psaraw4000}
\end{figure}

\begin{figure}
\includegraphics[width=\linewidth,height=\textheight,keepaspectratio]{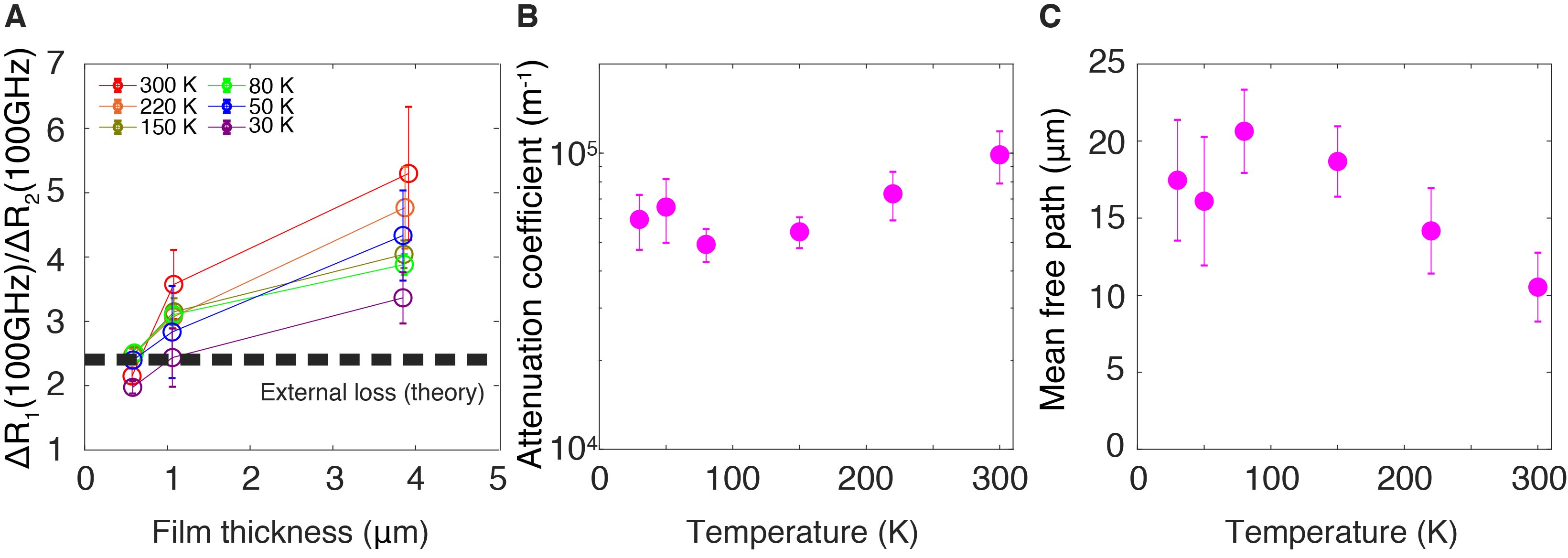}
\caption{(A) Ratio of the Fourier magnitude at 100 GHz measured at temperatures from 30 K to 300 K. The black dashed line indicates the predicted loss at the boundary between aSi and the sapphire from the AMM. Following Ref.~\cite{Hondongwa_PRB_2011}, external losses were experimentally deduced from the ratio at $\sim$ 500 nm. Internal damping was obtained from the values at $\sim 4$ $\mu$m using Eq.~\ref{eq:attn}. The film thicknesses were obtained from the time difference between the echoes. The error bar indicates standard deviation determined from multiple measurements. (B) Attenuation coefficients versus temperature using the 500 nm data for reflection loss and $\sim 4$ $\mu$m data for attenuation. (C) Mean free path versus temperature for the 100 GHz vibrations.}
\label{SfigaSi:psaanal}
\end{figure}

The amplitude signal versus the delay time for $\sim 500$ nm and $\sim 4$ $\mu$m samples are plotted in Figs.~\ref{SfigaSi:psaraw500} and \ref{SfigaSi:psaraw4000}. The resultant magnitude of the signal after one (two) round trip(s) are shown as $\Delta R_1 (t)$ ($\Delta R_2 (t)$). Following Ref.~\cite{Hondongwa_PRB_2011}, we analyze the data using Fourier analysis. First, we remove the background signal in $\Delta R_1 (t)$ and $\Delta R_2 (t)$. Then, $\Delta R_2 (t)$ is normalized by the maximum magnitude of $\Delta R_1 (t)$. Next, the signals are zero-padded to improve the frequency resolution of the fast Fourier transform (FFT). Finally, the time domain signals are windowed using a Hann window and the FFT is performed.
The resulting Fourier power spectra of each echo are plotted in Figs.~\ref{SfigaSi:psaraw500} and \ref{SfigaSi:psaraw4000}. The peak of the FFT occurs at $\sim 100$ GHz, indicating that 15 nm Al transducer can generate acoustic wave with a frequency $\sim 100$ GHz as expected. Despite a slight frequency dependence described in Ref.~\cite{Daly_PRB_2009}, following Refs.~\cite{Daly_PRB_2009, Hondongwa_PRB_2011, Morath_PRB_1996, Daly_PRB_2009}, rather than treating each Fourier amplitude separately, we compare Fourier amplitude at 100 GHz in Fig.~\ref{SfigaSi:psaanal}A. As in Ref.~\cite{Hondongwa_PRB_2011}, we find that the ratio of the Fourier magnitude at 100 GHz for 500 nm thickness is close to what is predicted by acoustic mismatch model (AMM) between aSi and the sapphire, $r = (\rho_{sapphire} v_{sapphire} - \rho_{aSi} v_{aSi}) / (\rho_{sapphire} v_{sapphire} +\rho_{aSi} v_{aSi}) =0.42 = 2.4^{-1} $, indicating that the loss for 500 nm thickness is dominated by transmission into the sapphire substrate. 



The attenuation coefficient for 100 GHz vibrations is obtained from
\begin{equation}
   \alpha = \frac{1}{d} \ln \left(\frac{r\Delta R_1 (\omega)}{\Delta R_2 (\omega)} \right)
\label{eq:attn}
\end{equation}
where $d$ is the round trip distance, $r$ is the inverse of the value from 500 nm aSi in Fig.~\ref{SfigaSi:psaanal}A, and $\Delta R_1 (\omega) / \Delta R_2 (\omega)$ is the value from the 4 $\mu$m thickness sample in Fig.~\ref{SfigaSi:psaanal}A.

The calculated attenuation coefficient is shown in Fig.~\ref{SfigaSi:psaanal}B. The measured coefficient is $\sim$ 10$^2$ - 10$^3$ $cm^{-1}$ with its maximum value at 300 K (987 $\pm$ 197 $cm^{-1}$). The measured value at room temperature is close to that reported from PSA in aSi (780 $\pm$ 160 $cm^{-1}$ )\cite{Hondongwa_PRB_2011} but substantially lower than that of vitreous silica ($\sim$ 10000 cm$^{-1}$ at $\sim$ 100 GHz) \cite{Klieber_APL_2011}. The mean free path of 100 GHz vibrations versus temperature is shown in Fig.~\ref{SfigaSi:psaanal}C. Our measured value is $\sim 10 - 20$ $\mu$m from 30 K ($\sim$ 17.5 $\pm$ 4 $\mu$m) to 300 K ($\sim$ 10.5 $\pm$ 2.2 $\mu$m) with a weak temperature dependence. 

\clearpage

\section{Crossover frequencies and $\kappa_{IR}$}

\begin{figure}[ht] \centering
\includegraphics[width=.8\linewidth,keepaspectratio]{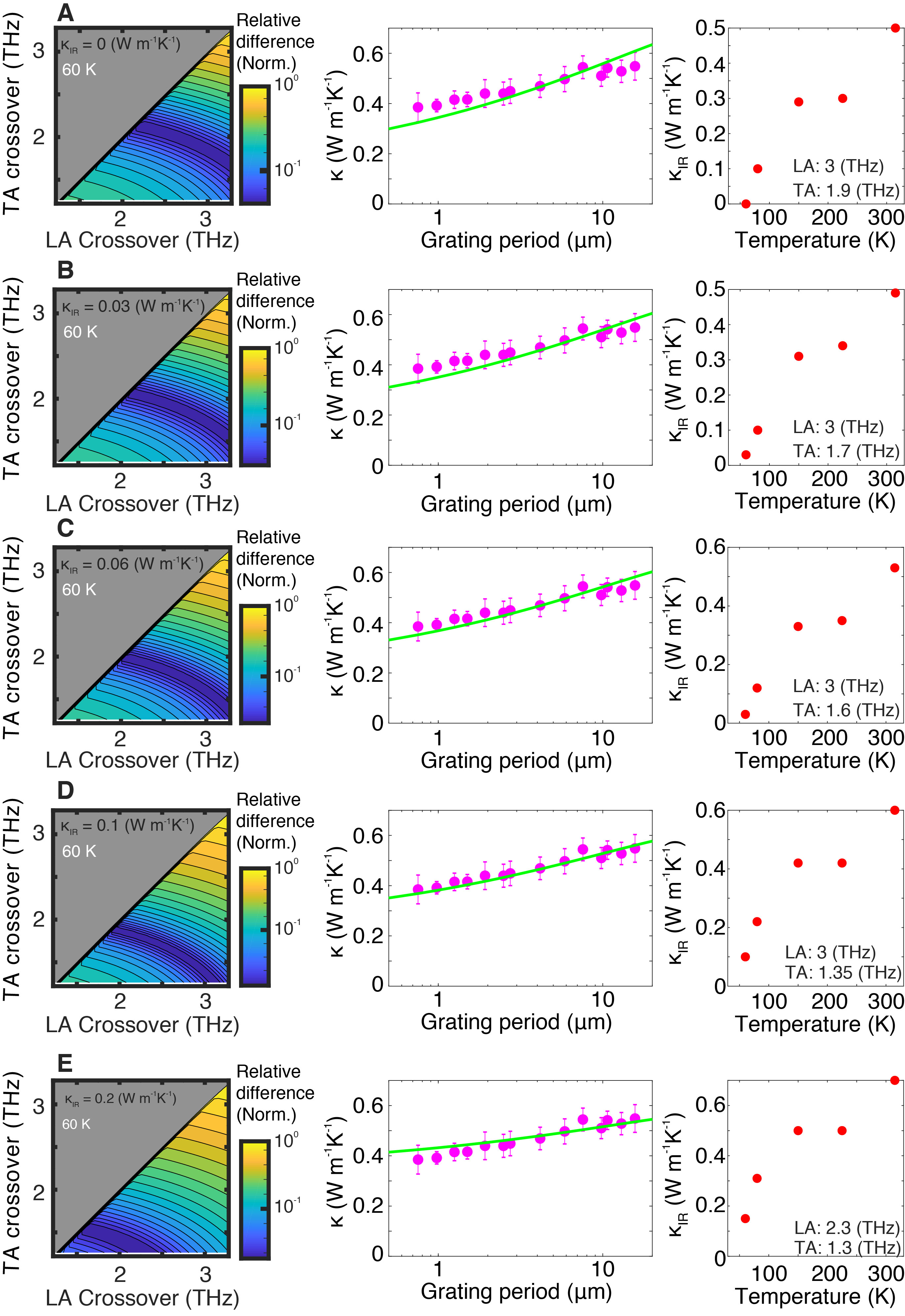}
\caption{(Left) Calculated relative differences at 60 K for (A) $\kappa_{IR}$ = 0 $W m^{-1}K^{-1}$, (B) $\kappa_{IR}$ = 0.03 $W m^{-1}K^{-1}$, (C) $\kappa_{IR}$ = 0.06 $W m^{-1}K^{-1}$, (D) $\kappa_{IR}$ = 0.1 $W m^{-1}K^{-1}$, (E) $\kappa_{IR}$ = 0.2 $W m^{-1}K^{-1}$. The blue area indicates the range of optimized LA / TA crossover frequency which minimize the relative difference. (Center) Calculated thermal conductivity versus grating period ($\kappa_{calc}$) using optimized crossover frequencies and experiments at 60 K. As the $\kappa_{IR}$ increases, the slope of the $\kappa_{calc}$ becomes less steep. (Right) Optimized $\kappa_{IR}$ at higher temperatures using optimized crossover frequency determined from minimum relative difference at 60 K.}
\label{SfigaSi:erranal}
\end{figure}

This section presents additional information regarding the values of the crossover frequencies and $\kappa_{IR}$ that are compatible with the data. The calculated relative differences between experiment $\kappa_{expt}$ and the calculation $\kappa_{calc}$ at 60 K using different $\kappa_{IR}$ values $\kappa_{IR} = 0 - 0.2$ $W m^{-1}K^{-1}$ are shown in the left of Fig.~\ref{SfigaSi:erranal}. Blue regions in the left of Fig.~\ref{SfigaSi:erranal} A - E indicate the cutoff frequencies that minimize the relative difference. We find that the optimized frequencies are in the range of 2-3 THz for LA and 1-2 THz for TA for all cases. As shown in the center column of Fig.~\ref{SfigaSi:erranal} A - E, $\kappa_{calc}$ at 60 K agrees with $\kappa_{expt}$ for $\kappa_{IR}$ $\leq$ 0.2 $W m^{-1}K^{-1}$ above which the trend of $\kappa_{calc}$ starts to deviate from that of $\kappa_{expt}$. The right column of Fig.~\ref{SfigaSi:erranal} shows optimized $\kappa_{IR}$ for 80 - 315 K using the optimized frequencies. The cutoff frequencies in the main text were selected to yield the overall best fit of thermal conductivity versus grating period at all temperatures. Therefore, although different choices for cutoff frequency and $\kappa_{IR}$ are possible, the frequency-dependence of the MFPs remains unchanged. A steep slope ($n=4$) in the sub-THz frequencies is necessary to explain to observed grating dependence of thermal conductivity.
\clearpage

\clearpage

\section{Self-heating and optical power}

In this section, we present an estimation of the steady and transient temperature rises that occurred during the experiment and justify the chosen optical powers. Care was taken to minimize the steady state and transient temperature rise of the sample while maintaining a signal magnitude of at least 300 - 400 $\mu$V.

\subsection{Temperature dependent optical properties in aSi}

Estimating the self-heating requires knowledge of the optical absorptance of the sample. We measured temperature-dependent reflectance and transmittance from which we estimated the absorptance of the amorphous silicon film. Table~\ref{tab:optpropaSi} shows the measured values. Accounting for optical reflections at various optical elements, the absorptance is around 50 - 55\% at the temperatures in this study. For simplicity we use 60\% for the calculations below.

\begin{table}[h] \centering
 \caption{\label{tab:optpropaSi}
 Optical properties of aSi measured at various temperatures.
 }

 \begin{tabular}{cccc}
 Cryostat temperature ($K$) & Reflectance (\%) & Transmittance (\%) & Absorptance (\%) \\ 
 \hline
 300 & 39 & 3 & 58\\
 210 & 35 & 4 & 62\\
 135 & 34 & 5 & 61\\
 60 & 32 & 5 & 63\\
  40 & 36 & 5 & 59\\
\end{tabular}

\end{table}

\clearpage


\subsection{Steady heating}
We estimate the steady heating due to the pump and probe pulses using a thermal resistor model in cylindrical coordinates. Consider the pump and probe beams as heat sources of radius $r_{pump}\equiv r_1$. The outer radius $r_2 \approx 500$ $\mu$m of the membrane is fixed at the cryostat temperature. The cylindrical conduction resistance is $R = \ln(r_2/r_1) /(2 \pi \kappa d)$ where $\kappa$ is the bulk thermal conductivity of the membrane. 
The temperature rise is then

\begin{eqnarray}
    \Delta T = P_{abs,avg} \frac{\ln(r_2/r_1)}{2 \pi \kappa d}
    \label{eq:steady}
\end{eqnarray}\\
where $P_{abs,avg}$ is the absorbed average power of the laser beam.
The absorbed average power was calculated using $P_{abs,avg} =\alpha P_{inc,avg} =\alpha P_{peak} D$ where P$_{inc,avg}$ is the incident average power on the sample, P$_{peak}$ is the peak power incident on the sample and $D$ is the duty cycle of the laser beam (1 for pump and 0.032 for probe from a chopper wheel).

Table~\ref{tab:steadyheat} shows the estimated steady temperature rise where $T_{h}$ is the cryostat temperature, $\kappa$ is the thermal conductivity used for calculation, $P_{pu,inc,avg}$ ($E_{pr,inc,avg}$) is the incident average power of the pump (probe) on the sample, $\Delta T_{pu}$ ($\Delta T_{pr}$) is the steady temperature rise due to pump (probe), and $\Delta T_{total}$ is the estimated total steady temperature rise of the sample, and $T_{calc}$ is the estimated temperature of the sample.

\begin{table}[h] \centering
 \caption{\label{tab:steadyheat}
Estimation of the steady heating of the sample.
 }

 \begin{tabular}{cccccccc}
$T_{h}$ ($K$) & $\kappa$ (Wm$^{-1}K^{-1}$) & $P_{pu,inc,avg}$ ($\mu$W) & $P_{pr,inc,avg}$ ($\mu$W) & $\Delta T_{pu}$ ($K$)  & $\Delta T_{pr}$ ($K$)& $\Delta T_{total}$ ($K$) & $T_{calc}$ ($K$)\\
\hline
 300 & 1.5 & 41 & 134 & 3.4 & 11.2 &  14.6 & $\sim 315$\\
 210& 1 & 27 & 88 & 3.4 & 11.0 & 14.4 & $\sim225$\\
 135 & 0.85 & 48 & 71 & 7.0 & 10.5 &  17.5 & $\sim 150$\\
  60 & 0.7 & 48 & 61& 8.5 & 10.9 & 19.4 &$\sim 80$\\
   40 & 0.55 & 44 & 46 & 10.1 & 10.5 & 20.6 & $\sim 60$\\
   \end{tabular}

\end{table}
\clearpage

\subsection{Transient heating}
We estimate the impulsive temperature rise induced by the absorbed optical energy. The temperature rise $\Delta T$ can be estimated from

\begin{eqnarray}
\Delta T = \frac{E_{abs}}{C_v V} \label{eq:Trans_HeatCond}
\end{eqnarray}\\
where $E_{abs}$ is absorbed laser energy, $V$ is the volume of the sample illuminated by the beam, and $C_{v}$ is the volumetric heat capacity \cite{Queen_PRL_2013}). The volume $V = \pi r^2 d$ where $d \approx 500$ nm is the estimated thickness of the film, and $r$ is the 1/$e^2$ beam radius (pump: 260 $\mu m$, and probe: 235 $\mu m$). 

The absorbed laser energy was calculated using $E_{abs} = \alpha E_{inc}$ where $\alpha$ is the measured absorptivity and $E_{inc}$ is the laser energy. The incident laser energy was obtained using
$E_{pu,inc} = P_{pu,inc} / f$ for pump and $E_{pr,inc} = P_{pr,inc} t_{chop}$ for probe where $E_{pu,inc}$ ($E_{pr,inc}$) is the pump (probe) energy incident on the sample, $P_{pu,inc}$ ($P_{pr,inc}$) is the incident pump (peak probe) power, $f$ is the repetition rate of the pump (200 Hz), and the $t_{chop}$ is the time duration of the probe (160 $\mu$s).

Table~\ref{tab:transheat} shows the estimated transient temperature rise where $T_{st}$ is the estimated temperature accounting for the steady temperature rise, $C_v$ is the heat capacity at $T = T_{st}$ in Ref.~\cite{Queen_PRL_2013}, $\Delta T_{pu}$ ($\Delta T_{pr}$) is the transient temperature rise due to pump (probe), and $\Delta T_{total}$ is the overall transient temperature rise of the sample.

\begin{table}[h] \centering
 \caption{\label{tab:transheat}
Estimation of the transient heating of the sample at several temperatures.
 }

 \begin{tabular}{ccccccc}

 $T_{st}$ (K) & $C_v$ (J$cm^{-3}$) & $E_{pu,inc}$ ($\mu$J) & $E_{pr,inc}$ ($\mu$J) & $\Delta T_{pu}$ (K)  & $\Delta T_{pr}$ (K)& $\Delta T_{total}$ (K)\\
 \hline
 315 &2.27 & 0.2 & 0.7 & 0.5& 2.0 & $\sim3$\\
 225& 2 & 0.1 & 0.4 & 0.4 & 1.5 & $\sim2$\\
 150 & 1.5 & 0.2 &  0.4 & 0.9 & 1.6 &$\sim3$\\
  80 & 0.73& 0.2 &  0.3 & 1.9 & 2.9 &$\sim5$\\
   60 & 0.52& 0.2 &  0.2 & 2.4 & 3.1 &$\sim6$\\
   \end{tabular}

\end{table}

\clearpage

\bibliographystyle{unsrtnat}
\bibliography{supplement}